\documentclass{elsart3p}

\usepackage{amssymb,amsmath}
\newcommand{\p}{\partial}

\begin{document}

\begin{frontmatter}



\title{From the Hubbard model to a systematic low-energy effective field theory for
  magnons and holes in an antiferromagnet}

\author[a]{Christoph Br\"ugger}
\author[a]{Florian K\"ampfer}
\author[a]{Markus Moser}
\author[b]{Michele Pepe}
\author[a]{Uwe-Jens Wiese}
\address[a]{Institute for Theoretical Physics, Bern University, Sidlerstrasse
  5, 3012 Bern, Switzerland}
\address[b]{Istituto Nazionale di Fisica Nucleare and
 Dipartimento di Fisica, Universit\`a di Milano-Bicocca, 3 Piazza della
 Scienza, 20126 Milano, Italy}

\begin{abstract}
The low-energy physics of antiferromagnets is governed by their Goldstone bosons --- the magnons --- and it is described by a low-energy effective field theory. In analogy to baryon chiral perturbation theory, we construct the effective field theory for magnons and holes in an antiferromagnet. It is a systematic low-energy expansion based on symmetry considerations and on the fact that the holes are located in pockets centered at $k=(\frac{\pi}{2a},\pm\frac{\pi}{2a})$.
Even though the symmetries are extracted from the Hubbard model, the effective theory is universal and makes model-independent predictions about the dynamical mechanisms in the antiferromagnetic phase. The low-energy effective theory has been used to investigate one-magnon exchange which leads to a $d$-wave-shaped bound state of holes.
\end{abstract}

\end{frontmatter}

\section{Introduction}
The investigation of the dynamics of holes in an antiferromagnet is a
challenging problem. Due to the strong correlation between the fermions in
these systems, a systematic study is very complicated. Away from half-filling,
microscopic systems such as the Hubbard or $t$-$J$ model, which may indeed
contain the relevant physics, are presently neither solved analytically nor numerically. Although there have been many attempts to understand high-temperature superconductors via their undoped precursors, the dynamical role of spin fluctuations remains a controversial issue.

In the following we will present a powerful and reliable tool for the
investigation of the dynamics in the antiferromagnetic phase. Inspired by 
chiral perturbation theory which is used to explore strongly interacting 
systems in particle physics, we have constructed a systematic low-energy 
effective field theory for holes and magnons. The effective theory is a systematic low-energy expansion that is based only on
symmetry considerations, basic principles of quantum field theory, and on the 
fact that the holes reside in pockets
centered at lattice momenta $(\frac{\pi}{2a},\pm \frac{\pi}{2a})$. The
construction of the effective theory is based on a specific microscopic model
Hamiltonian, but it is universally applicable to the entire class of
antiferromagnetic cuprates. Although the theory is not renormalizable, it
yields unambiguous results in the systematic low-energy expansion. In each
order of the expansion, the results depend only on a finite number of
material-specific low-energy parameters whose values are to be determined experimantally. It is very important to note that in contrast to the strongly correlated fermions of Hubbard-type models, the fermions of the effective field theory are quasi-particles that are weakly coupled to the magnons. Consequently, one may expect that the effective theory is more easily solvable than the underlying microscopic models.

\section{The effective Lagrangian}
In undoped antiferromagnets the global $SU(2)_s$ spin symmetry is
spontaneously broken down to $U(1)_s$. The low-energy physics of these
materials is thus governed by Goldstone bosons --- the magnons \cite{Cha89}, which are described by a unit-vector field 
$\vec e (x) = (e_1(x),e_2(x),e_3(x)) \in SU(2)_s/U(1)_s,$ with $\vec e(x)^2=1.$
Here $x=(x_1,x_2,t)$ is a point in continuum space-time. In order to couple
the magnons to the holes, we have constructed a nonlinear realization of the
$SU(2)_s$ symmetry from  the magnon field \cite{Kae05}. The global $SU(2)_s$ symmetry then manifests itself as a local $U(1)_s$ symmetry in the unbroken subgroup. This construction leads to an anti-Hermitean composite field
\begin{equation}
v_\mu(x)=iv_\mu^{a}(x)\sigma_a, \quad v_\mu^{\pm}(x)= v_\mu^{1}(x)\mp i v_\mu^{2}(x),
\end{equation}
where $\sigma_a$ are the Pauli matrices. The field $v_\mu(x)$ decomposes into
$v_\mu^{3}(x)$ which is an Abelian ``gauge" field and into $v_\mu^{\pm}(x)$
that represents two vector fields ``charged" under spin. These fields have a
clearly defined transformation behavior under the symmetries which the
effective theory shares with the microscopic models. The relevant symmetries
are the global $SU(2)_s$ symmetry, a $U(1)_Q$ fermion number symmetry together
with a set of discrete space-time symmetries such as displacement $D_i$ by one
lattice spacing in the $i$-direction, spatial rotation $O$ by 90 degrees,
reflection $R$ at a lattice axis, and time-reversal $T$.

Next we consider the holes, which live in momentum space pockets centered at
$k^{\alpha}=(\frac{\pi}{2a},\frac{\pi}{2a})$ and
$k^{\beta}=(\frac{\pi}{2a},-\frac{\pi}{2a})$. The holes are represented by
Grassmann fields $\psi_s^{f}(x)$, where the ``flavor" index $f=\alpha, \beta$
characterizes the corresponding hole pocket and the index $s=\pm$ denotes spin
parallel $(+)$ or antiparallel $(-)$ to the local staggered magnetization. The
construction of the low-energy effective theory is guided by symmetry
constraints. Analyzing the symmetry properties of a specific microscopic model
(the Hubbard model) and working out the commutation relations between the
various symmetries leads to a symmetry pattern that is  inherited by the
effective theory. Once the symmetry behavior of the hole and magnon fields is
determined, it is straightforward to construct the leading terms in the
effective Lagrangian. It depends only on a few real-valued low-energy
parameters, namely the spin stiffness $\rho_s$, the spinwave velocity $c$, the
rest mass and kinetic masses of a hole $M$, $M'$ and $M''$, a hole-one-magnon
coupling $\Lambda$, as well as on  hole-two-magnon couplings $N_1$ and
$N_2$. To leading order there are also three 4-fermion contact interactions
$G_1$, $G_2$, and $G_3$. Defining a $U(1)_s$ covariant derivative
\begin{equation}
D_\mu \psi_\pm^{f}(x)=[\p_\mu \pm i v_\mu^{3}(x)] \ \psi_\pm^{f}(x),
\end{equation}
the leading terms take the form
\begin{eqnarray}
&&\mathcal{L}[\psi^{f\dagger}_s,\psi^f_s,\vec e] =
\frac{\rho_s}{2} (\p_i \vec e \cdot \p_i \vec e + \frac{1}{c^2} \p_t \vec e\cdot  \p_t \vec e) \nonumber \\
&&+ \sum_{f, s } \ \{
M \psi^{f\dagger}_s \psi^f_s + \psi^{f\dagger}_s D_t \psi^f_s 
+\frac{1}{2 M'} D_i \psi^{f\dagger}_s D_i \psi^f_s \nonumber \\
&&+ \sigma_f \frac{1}{2 M''} (D_1 \psi^{f\dagger}_s D_2 \psi^f_s +
D_2 \psi^{f\dagger}_s D_1 \psi^f_s) \nonumber \\
&&+ \Lambda (\psi^{f\dagger}_s v^s_1 \psi^f_{-s}
+ \sigma_f \psi^{f\dagger}_s v^s_2 \psi^f_{-s}) +
N_1 \psi^{f\dagger}_s v^s_i v^{-s}_i \psi^f_s \nonumber \\
&&+ \sigma_f N_2 (\psi^{f\dagger}_s v^s_1 v^{-s}_2 \psi^f_s +
\psi^{f\dagger}_s v^s_2 v^{-s}_1 \psi^f_s) \nonumber \\
&&+\frac{G_1}{2} \psi^{f\dagger}_s \psi^f_s \psi^{f\dagger}_{-s} \psi^f_{-s}\}
\nonumber \\
&&+ \sum_{s}
\{G_2 \psi^{\alpha\dagger}_s \psi^\alpha_s \psi^{\beta\dagger}_s \psi^\beta_s +
G_3 \psi^{\alpha\dagger}_s \psi^\alpha_s
\psi^{\beta\dagger}_{-s} \psi^\beta_{-s}\}.
\end{eqnarray}
The sign $\sigma_f$ is $+$ for $f = \alpha$ and $-$ for $f = \beta$.

\section{Conclusions}
The above effective Lagrangian is a very powerful tool to investigate the
dynamics of holes in the antiferromagnetic phase. In \cite{Bru05} it has been
used to systematically derive the one-magnon exchange potential between
holes. It is shown that the attractive one-magnon exchange leads to bound
states of holes with $d$-wave symmetry. Furthermore, the effective Lagrangian
is presently used to study the dynamics of a doped system and the magnetic
spiral phase is explored. If spin-fluctuations remain among the relevant low-energy degrees of freedom, the effective theory may even be used to investigate the phenomenon of high-temperature superconductivity.

This work was supported in part by the SNF.

\end{document}